\documentclass[a4paper,twoside]{article}
\newcommand{\affil}[1]{$^{\rm #1}$}
\usepackage[authoryear]{natbib}
\bibpunct{(}{)}{;}{a}{}{,}
\usepackage{graphicx}
\usepackage{rotating} 
\date{} 
\def\mnras {{MNRAS}}
\def\apj {ApJ}
\def\apjs {ApJS}
\def\apjl {ApJL}
\def\aj {AJ}
\def\aap {A\&A}
\def\aaps {A\&AS}

\title{\large\bf\flushleft Are Dumbbell Brightest Cluster Members Signposts to Galaxy Cluster Activity?}
\author{\parbox{\textwidth}{\flushleft
\vspace{-0.5cm}
{\it K.\ A.\ Pimbblet\affil{A,B}}\\
\vspace{0.4cm}
{\small \affil{A}\,Department of Physics, University of Queensland, Brisbane, Queensland, QLD 4072, Australia}\\
{\small \affil{B}\,Email: pimbblet@physics.uq.edu.au}}}
\begin{document}
\twocolumn[
\begin{changemargin}{.8cm}{.5cm}
\begin{minipage}{.9\textwidth}
\vspace{-1cm}
\maketitle
%
%
\small{\bf Abstract:}
We assemble a sample of galaxy clusters whose brightest members
are dumbbell galaxies and compare them with a control sample in order
to investigate if they are the result of recent mergers.
We show that the dumbbell sample is no more likely than other
clusters to exhibit subclustering. 
However, they are much more likely to have at least one dumbbell
component possessing a significant peculiar velocity with respect 
to the parent cluster than a non-dumbbell
brightest cluster member.  
We interpret this in the context of seeing the clusters 
at various stages of post-merger relaxation.


\medskip
\medskip
\end{minipage}
\end{changemargin}
]
\small

\section{Introduction}
Brightest cluster member (BCM) 
galaxies have historically held a special 
place in both the theoretical and observational study of galaxy
evolution and formation since they are believed to be 
directly connected with the conditions in the cluster that 
they reside in.  
Consider, for instance, a BCM that is located at
the bottom of the gravitational well of a galaxy cluster: 
there, a process such as cannibalism (e.g.\ Ostriker \& Tremaine 1975)
would be at maximum efficiency there and we would 
be likely to observe BCMs with
multiple cores (e.g.\ Laine et al.\ 2003; 
Yamada et al.\ 2002; Oegerle \& Hill 2001;
Dubinski 1998; Lauer 1988; Hoessel \& Schneider 1985).  
However, if a more hierarchical paradigm is correct, then 
the BCM should have formed through multiple minor merger events
in a sub-group that was originally outside the cluster centre.
It would have then subsequently merged with the rest of the cluster 
(Merritt 1985), yielding several observational 
road-signs such as high peculiar velocities for the BCM 
and cluster substructure (see Woudt et al.\ 2008; 
Pimbblet et al.\ 2006; Oergle \& Hill 2001; Pinkney et al.\ 1996;
Quintana et al.\ 1996; Valentijn \& Casertano 1988).
 
Cases of dumbbell galaxies as BCMs 
(e.g.\ Gregorini et al.\ 1994; 1992)
are doubly interesting for these reasons -- 
these galaxies should be an indicator of 
the pre-virialization scenario and should therefore be accompanied 
by significant substructure indicative of recent cluster merger activity.  
Indeed, Quintana et al.\ (1996)
present evidence for a dumbbell system where the two
components of the BCM \emph{each} belong to major sub-groups that
are undergoing a merger (see also de Souza \& Quintana 1990).

What is unclear is whether dumbbell BCM clusters in general 
are special?
Does the presence of a dumbbell BCM point toward 
observable cluster activity 
such as subclustering (Quintana et al.\ 1996)
and large BCM peculiar velocities (Pimbblet et al.\ 2006)
as would be found during a merger event?  If so, then 
what stage is the merger event at?  
In order to make a pass at answering these questions, 
this work assembles a modest sample
of dumbbell BCMs from Gregorini et al.\ (1992) and a control
sample derived from the 2 degree field galaxy redshift survey
(2dFGRS; Colless et al.\ 2001).

The format of this paper is as follows.
In Section~2 we fully 
describe our derived dumbbell sample and control sample.
In Section~3, we ask whether clusters with dumbbell BCMs are 
more likely to have substructuring than any other clusters and then
in Section~4, we examine the incidence of significant peculiar
velocities in our samples are different.
We summarize our findings in Section~5.
Throughout this work, we adopt a cosmological concordance model with
values of $H_0 = 70$ kms$^{-1}$Mpc$^{-1}$, $\Omega_M = 0.3$ and
$\Omega_\Lambda = 0.7$.

\section{The Samples}
\begin{figure}[h]
\begin{center}
\includegraphics[scale=1, angle=0, width=3.in]{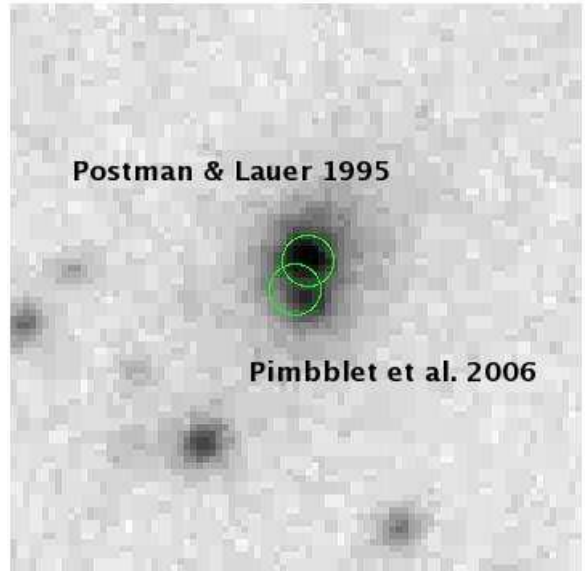}
\caption{Spectroscopic observations
of the components of the dumbbell BCM 
galaxy in Abell~3653 have been 
made by Postman \& Lauer (1995) and Pimbblet et al.\ (2006); 
the circles in the plot 
denote the location of where the redshift was taken.  
They give redshifts of 
$z=0.1091\pm0.0003$ and $z=0.1099\pm0.0002$ respectively.
Since both components are within $2\sigma$ of each other, we consider
this galaxy to be a confirmed dumbbell BCM 
galaxy for the purposes of this work.}\label{fig:a3653}
\end{center}
\end{figure}

The Gregorini et al.\ (1992) sample claims to be a 
volume limited \& homogeneous 
sample of dumbbell BCM galaxies which should be ideal for our
purposes.  It is 
based on the Abell catalogue (Abell 1958; Abell et al.\ 1989)
and is complete out to a comoving distance
of 210h$^{-1}$ Mpc (Gregorini et al.\ 1992).
In addition, they state 
there are no possible selection biases present in this sample that
would have prevented the successful detection and
identification of dumbbell galaxies for this
sample (Gregorini et al.\ 1992; see also 
Scaramella et al.\ 1991). 
There are a couple of 
debatable clusters that 
could have been added to this sample 
(e.g.\ Abell~3323; Gregorini et al.\ 1994), but were left out.
We do not view the exclusion of these clusters as having a significant
impact on the ensemble.
However, our review of the available 
literature redshifts demonstrates that this sample
does contain several clusters with $z>0.1$ (Table~1) 
due to earlier cluster redshifts being under-estimated.
We do not regard this as a major impediment to the present
investigation either, since the time evolved between $z\sim0.07$ and 
$z\sim0.10$ is much smaller (i.e.\ $<0.5$ Gyr) than the 
expected time it would take for sub-clusters to fully merge
(Lacey \& Cole 1993).
For this work, we restrict ourselves to the complete sample of 
dumbbell galaxies -- Table~1 from Gregorini et al.\ (1992; 
herein the dumbbell sample) -- 
with one exception: Abell~3653.  Abell~3653 is listed by 
Gregorini et al.\ (1992) as only a `possible' dumbbell BCM.
However, a combination of 
observations made by Postman \& Lauer (1995) and 
Pimbblet et al.\ (2006) show that this cluster should also be
considered to be a confirmed dumbbell BCM galaxy 
(Figure~\ref{fig:a3653}) and we include it in the present
dumbbell sample.

Redshift information for the dumbbell sample is initially obtained by
downloading all redshifts within 30 arcmin of each cluster
from the NASA Extra-galactic Database (NED).  At $z\sim0.1$, this
corresponds to a radius of 1.6 Mpc from the cluster centre.  
Several clusters in the resultant dumbbell sample 
(Abell~2824, 3098, 3368, 3397, \& 3740) generate very few members 
($N<20$) within 30 arcmin 
($\approx$ an Abell radius)
and are eliminated from the final
sample at this stage since it is
likely that any subclustering or cluster velocity 
dispersion measurement
would be unreliable (Girardi et al.\ 1993).  
For the remaining sample, we compute the
mean cluster velocity and velocity dispersion ($\sigma$) using
the `gapping' procedure of Zabludoff et al.\ (1990; 1993) and
tabulate these values in Table~1.
The final sample yields 13 dumbbell 
BCM galaxy clusters for us to work with that
have a range of velocity dispersions consistent with large 
and massive clusters (cf.\ Ebeling et al.\ 2007).

Our control sample is obtained from the 2dFGRS
cluster sample of De Propris et al.\ (2002; see Colless et al.\ 2001
for a description of this survey).  
This is essentially a complete catalogue at $cz< \ \sim
35000$ km s$^{-1}$ of 
Abell (Abell 1958; Abell et al.\ 1989) clusters. 
In order to attempt to match the relatively high velocity 
dispersions in the dumbbell sample, we restrict the 
control sample to only those clusters with a high X-ray 
luminosity ($L_X>0.5 \times 10^{44}$ erg s$^{-1}$; 
Ebeling et al.\ 1996; Cruddace et al.\ 2002).  We also trim from
this sample any clusters with poor completeness levels 
(i.e.\ less than 20 galaxies within an Abell radius of the cluster centre).
Our final control sample consists of 14 clusters and 
is presented in Table~2 with mean velocities and velocity
dispersions sourced from De Propris et al.\ (2002).  
Although the control sample has a smaller absolute range of
$\sigma$ (597--1038 kms$^{-1}$) than the dumbbell cluster
sample (455--1376 kms$^{-1}$), the median values
(783.5 kms$^{-1}$ and 825 kms$^{-1}$ respectively)
are very similar.  
The average redshift of the control sample 
($cz=26010$ kms$^{-1}$)
is also somewhat higher
than the dumbbell sample ($cz=18854$ kms$^{-1}$).  In look-back
time this is barely 0.3 Gyr, so again we do not regard this as 
significant for this work 
due to the comparative time taken for clusters to violently relax
after a merger event (Lacey \& Cole 1993).
With the control sample being sourced from 2dFGRS, we ensure that
all the constituent galaxies are very homogeneously sampled
down to $b_J=19.45$ (Colless et al.\ 2001); albeit not at 
100\% completeness levels (see De Propris et al.\ 2002).  
Conversely, our dumbbell sample
is a collection of diverse redshifts from multiple 
sources each with different purposes and 
associated selection limits imposed (see Table~1).  
We also eyeball the BCMs in the control sample to ensure that
they are not dumbbells themselves that are outside the Gregorini
et al.\ (1992) selection limits. Only two of them give us cause for
concern: Abell~2811 has a nearby faint companion, 
but would not be considered a dumbbell in the definition of Gregorini
et al.\ (1992); Abell~S1136 has multiple galaxies near to the BCM
that may be interacting with it, but again it is not a dumbbell
and its removal from the control sample does not affect our conclusions.

%
%
\begin{sidewaystable*}
\begin{center}
\caption{The dumbbell cluster sample$^{a}$.  \hfil}
\begin{tabular}{llllllllllllll}
\hline
Cluster &  RA      & Dec     & N(30')$^{b}$ & $cz$         & $\sigma$     & $R_{V}$ & N$_{R_V}$ & $\Delta_{R_V}$ &  P($\Delta$)$_{R_V}$ & N$_{2R_V}$ & $\Delta_{2R_V}$ &  P($\Delta$)$_{2R_V}$\\
        &  (J2000) & (J2000) &   & (kms$^{-1}$) &  (kms$^{-1}$) & (Mpc)   &     &     &  \\    
\hline
Abell~2860 & 1$^{h}$4$^{m}$20.60$^{s}$ & -39$^{o}$2961.4" & 59 & 31738$\pm$97  & $741^{+80}_{-60}$ & 1.48 & 28 & 76 & 0.990 & 45 & 148 & 0.985 \\
Abell~2911 & 1$^{h}$26$^{m}$12.87$^{s}$ & -37$^{o}$3381.8" & 50 & 24110$\pm$102 & $720^{+85}_{-63}$ & 1.44 & 34 & 30 & 0.720 & 50 & 41 & 0.959 \\
Abell~3151 & 3$^{h}$40$^{m}$27.73$^{s}$ & -28$^{o}$2543.0" & 68 & 19992$\pm$124 & $1013^{+101}_{-78}$ & 2.03 & 60 & 115 & 0.009 & 103 & 230 & 0.016 \\
Abell~3266 & 4$^{h}$31$^{m}$11.92$^{s}$ & -61$^{o}$1462.6" & 286 & 17852$\pm$81 & $1376^{+61}_{-54}$ & 2.75 & 248 & 277 & 0.313 & 399 & 524 & $<$0.001 \\
Abell~0533 & 5$^{h}$1$^{m}$30.79$^{s}$ & -22$^{o}$2202.4" & 27 & 13932$\pm$229 & $1191^{+208}_{-136}$ & 2.38 & 35 & 92 & $<$0.001 & 52 & 146 & $<$0.001 \\
Abell~3391 & 6$^{h}$26$^{m}$15.43$^{s}$ & -53$^{o}$2452.3" & 104 & 16070$\pm$127 & $1294^{+101}_{-82}$ & 2.59 & 131 & 207 & $<$0.001 & 241 & 440 & $<$0.001 \\
Abell~3528 & 12$^{h}$54$^{m}$18.36$^{s}$ & -29$^{o}$75.6" & 146 & 16219$\pm$87 & $1047^{+67}_{-56}$ & 2.09 & 145 & 194 & 0.017 & 236 & 327 & $<$0.001 \\
Abell~3532 & 12$^{h}$57$^{m}$19.08$^{s}$ & -30$^{o}$1332.7" & 167 & 16504$\pm$60 & $780^{+47}_{-40}$ & 1.56 & 117 & 150 & 0.061 & 233 & 303 & 0.018 \\
Abell~3535 & 12$^{h}$57$^{m}$48.60$^{s}$ & -28$^{o}$1752.1" & 55 & 20190$\pm$61 & $455^{+51}_{-38}$ & 0.91 & 37 & 46 & 0.041 & 53 & 78 & 0.010 \\
Abell~3570 & 13$^{h}$46$^{m}$50.88$^{s}$ & -37$^{o}$3298.3" & 31 & 11261$\pm$83 & $463^{+74}_{-50}$ & 0.93 & 19 & 17 & 0.589 & 37 & 34 & 0.627 \\
Abell~3653 & 19$^{h}$52$^{m}$37.92$^{s}$ & -52$^{o}$74.1" & 106 & 32234$\pm$81 & $836^{+64}_{-52}$ & 1.67 & 50 & 61 & 0.164 & 92 & 123 & 0.044 \\
Abell~3716 & 20$^{h}$51$^{m}$16.56$^{s}$ & -52$^{o}$2503.4" & 124 & 13517$\pm$74 & $825^{+58}_{-48}$ & 1.65 & 110 & 113 & 0.226 & 196 & 365 & $<$0.001 \\
Abell~3744 & 21$^{h}$7$^{m}$13.80$^{s}$ & -25$^{o}$1733.7" & 69 & 11483$\pm$72 & $596^{+59}_{-45}$ & 1.19 & 64 & 88 & 0.066 & 88 & 129 & 0.007 \\
\hline
\end{tabular}
\ \\
$^{a}$Redshifts for this sample are sourced from: Fairall 1988;
Teague et al.\ 1990;
da Costa et al.\ 1991;
Dalton et al.\ 1994;
Collins et al.\ 1995; Postman \& Lauer 1995; Quintana et al.\ 1995; 
Loveday et al.\ 1996; Quintana \& Ramirez 1996; Shectman et al.\ 1996; 
Caldwell \& Rose 1997; 
da Costa et al.\ 1998; Katgert et al.\ 1998; Vettolani et al.\ 1998; 
Wegner et al.\ 1999; 
Schindler 2000;
Bardelli et al.\ 2001; Colless et al.\ 2001; Donnelly et al.\ 2001;
Christlein \& Zabludoff 2003; Kaldare et al.\ 2003; Paturel et al.\ 2003; 
Jones et al.\ 2004; Smith et al.\ 2004; 
Pimbblet et al.\ 2006.\\
$^{b}$This is the number of NED galaxies within 30 arcmin used to compute $cz$ and $\sigma$ from.\\
  \label{tab:dumbbells}
\end{center}
\end{sidewaystable*}

\section{Substructure}
In order to place the clusters onto a common scale
and perform a meaningful comparison of subclustering, we
use a virial radius estimator
derived by Girardi et al.\ (1998): 
$R_v = 0.002 \sigma h^{-1}_{100}$ Mpc.
Cluster members are then defined to be those galaxies
whose velocity is within $3\sigma$ of the cluster velocity.
We add the caveat that although this approximation's 
validity for semi-virialized systems may not be ideal, it 
is sufficient for our purposes of placing the clusters 
on to a $\sim$common scale.

For the cluster centres, we use the quoted NED cluster 
centres -- this choice of cluster centre is somewhat arbitrary, 
but it will not dramatically affect the final subclustering 
result since we are sampling the cluster members from a relatively 
large radius away from this centre.
The resultant number of galaxies and values for $R_v$ 
for each of the clusters are given in Tables~1 \&~2.

There are a number of statistical tools available to
assess the degree of subclustering in each cluster, ranging from
2 dimensional searches for asymmetry (e.g.\ West et al.\ 1988)
and bimodality (e.g.\ Fitchett \& Webster 1987) to more complex 3 dimensional
tests (e.g. Dressler \& Shectman 1988; among others).
Given all these different tools, Pinkney et al.\ (1996) made
extensive tests to determine the relative merits of these tools and 
concluded that the Dressler \& Shectman (1988) $\Delta$ test is
the best one to use to find substructure in arbitrary cases.  
Its only real limitation is an insensitivity to equal mass
mergers in the plane of the sky and superpositions of sub-groups
(Pinkney et al.\ 1996).  Both of these situations would require
special and unusual lines of sight to the cluster and are therefore
considered to be rare events. 
Importantly for this study, the DS test 
will be able to detect substructure in 
3:1 mergers with reasonable confidence (circa 95\%) 
down to a sample size of even 30 galaxies (Fig.~27 in 
Pinkney et al.\ 1996).  At a sample size of $>$60, this confidence
rapidly grows to $>$99\%.  

We therefore proceed by applying the Dressler \& Shectman (1988) 
approach to each of our clusters and we refer the reader to that 
publication for details of its algorithmic execution.  For the purposes
of this work, it is sufficient to note that if there is little or no
substructure, then we may expect the resultant $\Delta$ statistic to
be of approximately the same value as the number of galaxies sampled. 
The final parameter of merit is then $P(\Delta)$ which gives the
probability of the observed value of $\Delta$ occurring randomly when
the redshifts of cluster members are randomly assigned to other members
in a Monte Carlo fashion (i.e. very low values of $P(\Delta)$ indicate
the presence of substructure).

Values of $\Delta$ and $P(\Delta)$ are computed for each cluster 
within radial limits of both $R_V$ and $2R_V$ (Tables~1 \& 2).
The reason for looking at both these radii is that it may be possible
that the $\Delta$ statistic is insensitive to 
substructure at small ($<2$Mpc) radii (Pinkney et al.\ 1996).
We also plot the `bubble plots' in Figure~2 for the results of
the DS test for the dumbbell sample.  
In these plots, substruture can be interpreted as 
spatially close overlapping circles (Dressler \& Shectman 1988).
Within $R_V$, we find that the fraction of clusters with certain
substructure (i.e.\ $P(\Delta)<0.001$) is 2/13 (15\%) for the
dumbbell sample and 1/14 (7\%) for the control sample.  
At $2R_V$, these fractions change to 5/13 (39\%) and 4/14 (29\%) 
respectively  (in addition, a further two of control clusters 
are very close to being considered as having subclustering).
We consider the difference in subclustering fractions between
the two cluster samples to be statistically insignificant.

%
%
\begin{sidewaystable*}
\begin{center}
\caption{The control cluster sample.  \hfil}
\begin{tabular}{lllllllllllll}
\hline
Cluster &  RA      & Dec     &  $cz$         & $\sigma$     & $R_{V}$ & N$_{R_V}$ & $\Delta_{R_V}$ &  P($\Delta$)$_{R_V}$ & N$_{2R_V}$ & $\Delta_{2R_V}$ &  P($\Delta$)$_{2R_V}$ \\
        &  (J2000) & (J2000)   & (kms$^{-1}$) &  (kms$^{-1}$) & (Mpc)   &     &     &  \\    
\hline
Abell 2734 & 0$^{h}$11$^{m}$19.44$^{s}$ & -28$^{o}$3099.6" & $18646 \pm 94$  & $1038^{+73}_{-84}$ & 2.08 & 120 & 153 & 0.127 & 214 & 427 & $<$0.001 \\
Abell 2751 & 0$^{h}$16$^{m}$19.92$^{s}$ & -31$^{o}$1314.0" & $31863 \pm 103$ & $763^{+80}_{-95}$ & 1.47 & 43 & 67 & 0.014 & 99 & 128 & 0.027 \\
Abell 2755 & 0$^{h}$17$^{m}$34.80$^{s}$ & -35$^{o}$669.6" & $28469 \pm 135$ & $829^{+102}_{-129}$ & 1.66 & 29 & 21 & 0.844 & 64 & 87 & 0.180 \\
Abell 2798 & 0$^{h}$37$^{m}$27.12$^{s}$ & -28$^{o}$1886.4" & $33516 \pm 108$ & $739^{+84}_{-102}$ & 1.48 & 41 & 60 & 0.008 & 77 & 120 & 0.004 \\
Abell 2811 & 0$^{h}$42$^{m}$8.88$^{s}$ & -28$^{o}$1933.2" & $32557 \pm 168$ & $988^{+126}_{-162}$ & 1.98 & 83 & 111 & 0.193 & 175 & 264 & $<$0.001 \\
Abell 2829 & 0$^{h}$51$^{m}$19.20$^{s}$ & -28$^{o}$1868.4" & $33565 \pm 123$ & $793^{+94}_{-117}$ & 1.59 & 39 & 50 & 0.043 & 69 & 118 & 0.010 \\
Abell 3027 & 2$^{h}$30$^{m}$51.12$^{s}$ & -33$^{o}$349.2" & $23166 \pm 97$  & $907^{+76}_{-88}$ & 1.81 & 85 & 111 & 0.041 & 166 & 233 & 0.002 \\
Abell 0389 & 2$^{h}$51$^{m}$28.80$^{s}$ & -24$^{o}$3416.4" & $34085 \pm 118$ & $667^{+90}_{-115}$ & 1.33 & 26 & 24 & 0.632 & 45 & 55 & 0.173 \\
Abell 3094 & 3$^{h}$11$^{m}$25.18$^{s}$ & -26$^{o}$3240.0" & $20475 \pm 77$  & $774^{+61}_{-70}$ & 1.55 & 89 & 110 & 0.084 & 140 & 233 & $<$0.001 \\
Abell 0957 & 10$^{h}$13$^{m}$49.80$^{s}$ & -0$^{o}$3254.4" & $13623 \pm 79$  & $722^{+63}_{-73}$  & 1.44 & 77 & 95 & 0.084 & 115 & 149 & 0.065 \\
Abell 1651 & 12$^{h}$59$^{m}$24.00$^{s}$ & -4$^{o}$680.4" & $25152 \pm 107$ & $817^{+83}_{-99}$  & 1.63 & 90 & 127 & 0.194 & 173 & 218 & 0.213 \\
Abell 1750 & 13$^{h}$30$^{m}$49.68$^{s}$ & -1$^{o}$3110.4" & $25647 \pm 113$ & $981^{+88}_{-104}$ & 1.96 & 101 & 168 & $<$0.001 & 193 & 296 & $<$0.001 \\
Abell 2597 & 23$^{h}$25$^{m}$16.68$^{s}$ & -12$^{o}$446.4" & $24691 \pm 117$ & $597^{+117}_{-90}$ & 1.19 & 21 & 16 & 0.529 & 36 & 37 & 0.276 \\
Abell S1136 & 23$^{h}$36$^{m}$14.76$^{s}$ & -31$^{o}$2156.4" & $18688 \pm 92$ & $617^{+72}_{-87}$ & 1.23 & 38 & 51 & 0.097 & 55 & 79 & 0.025 \\
\hline
\end{tabular}
\ \\
  \label{tab:control}
\end{center}
\end{sidewaystable*}

This is somewhat surprising as we may have 
naively expected there
to have been recent cluster activity in these dumbbell cluster 
systems (cf.\ Quintana et al.\ 1996) compared to the general
cluster population.  
We re-emphasize the caveat of the DS test limitations: with some
of our clusters (size$\sim30$), 
the confidence level in the $\Delta$ statistic may
drop to only 95\% to be able to detect a 3:1 merger 
(Pinkney et al.\ 1996) -- $\sim$5 clusters
at radii$<R_V$ and $\sim1$ cluster at $<2R_V$ in the dumbbell sample; 
and similar numbers in the control sample.  
Even with this confidence level and sample sizes, the data suggest
that the incidence of 3:1 mass mergers is equal in both samples.
Given that Quintana et al.\ (1996) demonstrate
that each component of the dumbbell in NGC 4782/3 system occupies
different sub-groups, it is therefore a 
natural next step to ask whether the other galaxies in 
our samples also have large peculiar velocities?

\section{Peculiar Velocities}
Pimbblet et al.\ (2006) have already noted that the 
dumbbell BCM galaxy from Abell~3653 has 
a very large peculiar velocity relative to the cluster frame 
(for both components of the dumbbell -- Fig~\ref{fig:a3653}).  
Using the positions of the individual dumbbell components given
by Gregorini et al.\ (1994), we compare each of their redshifts 
with the mean cluster velocity from Table~1. The results of 
this are displayed in Table~3. The error listed on the 
velocity offset, $\Delta cz$, in this Table is simply 
the mean cluster velocity error (Table~1)
added in quadrature to the galaxy velocity error given by 
NED (where more than 1 redshift measurement of a given 
dumbbell component is available, we choose the most recent since
there is some uncertainty concerning which redshift may
refer to what component in older measurements; e.g.\ Green et al.\ 1988).
We perform a similar analysis on the control sample, 
identifying the BCM by eye and a magnitudinal comparison using 
NED (see Table~4).  

For the control sample, we find that 3/14 (21\%) of the clusters
have BCMs with significant peculiar velocities 
(i.e.\ $>3$ standard errors away from $| \Delta cz | = 0$).
Meanwhile in the dumbbell sample, 8/13 (62\%) clusters 
have at least 1 dumbbell
component with significant peculiar velocities (7 of which are 
large, $>300$kms$^{-1}$); 
but consistent to the control sample, 
only 3/13 (23\%) have \emph{both} dumbbell components with 
significant peculiar velocities.
We note that for Abell~0533 the two dumbbell components
have a very significant velocity offset ($>1600$ kms$^{-1}$!)
which questions whether this system should be considered
as a proper dumbbell system.  
Notwithstanding Abell~0533, we suggest that a 
cluster with a BCM dumbbell system
is more likely to possess a significant peculiar velocity
in at least one of the dumbbell components 
than a non-dumbbell BCM cluster.  
Save for Abell~0533, no peculiar velocity is beyond 1.1 times the
velocity dispersion measurement of the clusters, very much 
inline with
the results of Valentijn \& Casertano (1988).

%
%
\begin{table*}
\begin{center}
\caption{Peculiar velocities of the dumbbells.  \hfil}
\begin{tabular}{llllllllll}
\noalign{\medskip}
\hline
Dumbbell    & RA       & Dec     & $| \Delta cz |$  & Significance \\
Component   & (J2000)  & (J2000) & (kms$^{-1}$) \\
\hline
Abell~2860 1      & 1$^{h}$4$^{m}$50.13$^{s}$ & -39$^{o}$2749.6" & 539$\pm$110 & 4.90 \\
~~''~~~~~~''~~~ 2 & 1$^{h}$4$^{m}$50.11$^{s}$ & -39$^{o}$2749.8" & 634$\pm$(97)$^{a}$ & (6.54)$^{a}$  \\
Abell~2911 1      & 1$^{h}$26$^{m}$5.42$^{s}$ & -37$^{o}$3472.9" & 143$\pm$120 & 1.19 \\
~~''~~~~~~''~~~ 2 & uncertain & uncertain & uncertain$^{b}$ & -- \\
Abell~3151 1      & 3$^{h}$40$^{m}$26.94$^{s}$ & -28$^{o}$2437.5" & 376$\pm$132 & 2.85 \\
~~''~~~~~~''~~~ 2 & 3$^{h}$40$^{m}$25.14$^{s}$ & -28$^{o}$2439.0" & 426$\pm$128 & 3.33 \\
Abell~3266 1      & 4$^{h}$31$^{m}$13.29$^{s}$ & -61$^{o}$1631.8" & 247$\pm$103 & 2.40 \\
~~''~~~~~~''~~~ 2 & 4$^{h}$31$^{m}$12.18$^{s}$ & -61$^{o}$1635.1" & 114$\pm$101 & 1.13 \\
Abell~0533 1      & 5$^{h}$1$^{m}$8.29$^{s}$ & -22$^{o}$2095.5" & 255$\pm$241 & 1.06 \\
~~''~~~~~~''~~~ 2 & 5$^{h}$1$^{m}$6.63$^{s}$ & -22$^{o}$2096.2" & 1906$\pm$230 & 8.29 \\
Abell~3391 1      & 6$^{h}$26$^{m}$20.22$^{s}$ & -53$^{o}$2494.0" & 489$\pm$133 & 3.68 \\
~~''~~~~~~''~~~ 2 & 6$^{h}$26$^{m}$17.80$^{s}$ & -53$^{o}$2489.4" & 68$\pm$142 & 0.48 \\
Abell~3528 1      & 12$^{h}$54$^{m}$23.40$^{s}$ & -29$^{o}$64.8" & 100$\pm$107 & 0.93 \\
~~''~~~~~~''~~~ 2 & 12$^{h}$54$^{m}$22.32$^{s}$ & -29$^{o}$46.8" & 0$\pm$101 & 0.00 \\
Abell~3532 1      & 12$^{h}$57$^{m}$21.96$^{s}$ & -30$^{o}$1308.9" & 257$\pm$64 & 4.02 \\
~~''~~~~~~''~~~ 2 & 12$^{h}$57$^{m}$19.80$^{s}$ & -30$^{o}$1312.9" & 36$\pm$67 & 0.54 \\
Abell~3535 1      & 12$^{h}$57$^{m}$55.44$^{s}$ & -28$^{o}$1720.0" & 485$\pm$92 & 5.27 \\
~~''~~~~~~''~~~ 2 & 12$^{h}$57$^{m}$54.72$^{s}$ & -28$^{o}$1728.0" & 462$\pm$75 & 6.16 \\
Abell~3570 1      & 13$^{h}$46$^{m}$47.28$^{s}$ & -37$^{o}$3268.4" & 24$\pm$111 & 0.22 \\
~~''~~~~~~''~~~ 2 & 13$^{h}$46$^{m}$46.92$^{s}$ & -37$^{o}$3282.4" & 49$\pm$120 & 0.41 \\
Abell~3653 1      & 19$^{h}$53$^{m}$3.48$^{s}$ & -52$^{o}$133.2" & 736$\pm$105 & 7.01 \\
~~''~~~~~~''~~~ 2 & 19$^{h}$53$^{m}$2.76$^{s}$ & -52$^{o}$134.6" & 495$\pm$126 & 3.93 \\
Abell~3716 1      & 20$^{h}$52$^{m}$0.48$^{s}$ & -52$^{o}$2718.0" & 559$\pm$92 & 6.08 \\
~~''~~~~~~''~~~ 2 & 20$^{h}$51$^{m}$56.88$^{s}$ & -52$^{o}$2710.8" & 255$\pm$88 & 2.90 \\
Abell~3744 1      & 21$^{h}$7$^{m}$25.68$^{s}$ & -25$^{o}$1543.3" & 27$\pm$94 & 0.29 \\
~~''~~~~~~''~~~ 2 & 21$^{h}$7$^{m}$24.60$^{s}$ & -25$^{o}$1557.0" & 211$\pm$75 & 2.81 \\
\hline
\noalign{\smallskip}
\end{tabular}
\ \\
$^{a}$Redshift error not recorded for this component -- 
the error quoted is simply the error on the cluster mean 
velocity and should therefore be
taken as a lower bound.\\
$^{b}$Unable to unambiguously distinguish the second
component and its redshift from the first component.
\\
  \label{tab:pecvdumbbells}
\end{center}
\end{table*}

%
%
\begin{table*}
\begin{center}
\caption{BCM peculiar velocities in the control sample.  \hfil}
\begin{tabular}{lllllllllllllll}
\noalign{\medskip}
\hline
Cluster & BCM RA  & BCM Dec & $| \Delta cz |$ & Significance \\
        & (J2000) & (J2000) & (kms$^{-1}$) \\
\hline
Abell 2734 & 0$^{h}$11$^{m}$21.67$^{s}$ & -28$^{o}$3075.8" & 171$\pm$101 & 1.69\\
Abell 2751 & 0$^{h}$16$^{m}$13.65$^{s}$ & -31$^{o}$1392.1" & 245$\pm$261 & 0.94 \\
Abell 2755 & 0$^{h}$17$^{m}$40.99$^{s}$ & -35$^{o}$720.7" & 284$\pm$224 & 1.27 \\
Abell 2798 & 0$^{h}$37$^{m}$32.20$^{s}$ & -28$^{o}$1915.9" & 249$\pm$116 & 2.15 \\
Abell 2811 & 0$^{h}$42$^{m}$8.83$^{s}$ & -28$^{o}$1928.8" & 23$\pm$179 & 0.13 \\
Abell 2829 & 0$^{h}$51$^{m}$22.53$^{s}$ & -28$^{o}$1897.5" & 416$\pm$126 & 3.30 \\
Abell 3027 & 2$^{h}$30$^{m}$49.42$^{s}$ & -33$^{o}$371.8" & 384$\pm$132 & 2.91 \\
Abell 0389 & 2$^{h}$51$^{m}$24.80$^{s}$ & -24$^{o}$3399.4" & 331$\pm$191 & 1.73 \\
Abell 3094 & 3$^{h}$11$^{m}$25.00$^{s}$ & -26$^{o}$3351.9" & 75$\pm$92 & 0.82 \\
Abell 0957 & 10$^{h}$13$^{m}$38.28$^{s}$ & -0$^{o}$3331.8" & 253$\pm$95 & 2.66 \\
Abell 1651 & 12$^{h}$59$^{m}$22.56$^{s}$ & -4$^{o}$706.3" & 419$\pm$111 & 3.97 \\
Abell 1750 & 13$^{h}$30$^{m}$50.76$^{s}$ & -1$^{o}$3097.0" & 441$\pm$116 & 3.80 \\
Abell 2597 & 23$^{h}$25$^{m}$19.92$^{s}$ & -12$^{o}$446.0" & 189$\pm$118 & 1.60 \\
Abell S1136 & 23$^{h}$36$^{m}$16.56$^{s}$ & -31$^{o}$2169.3" & 19$\pm$112 & 0.17 \\
\hline
\noalign{\smallskip}
\end{tabular}
  \label{tab:pecvcontrol}
\end{center}
\end{table*}

\section{Discussion and Conclusions}
Even though the dumbbell sample is a collection of 
redshifts from many disparate sources, we believe that 
we have assembled a sufficient quantity of (bright) cluster
members to adequately map out each of the clusters to a similar
level found in 2dFGRS (cf.\ Tables~1 \&~2).  
Although the sample is likely not an optimal (homogeneously selected)
one, we are nonetheless confident that the results presented 
would not change significantly with the addition of further
observations.  Indeed, subtracting a small 
percentage of galaxies from 
the better sampled clusters produces no significant change in 
the measured peculiar velocities, or substructuring.  

Our results show that dumbbell BCM clusters are no more or less
likely than other clusters to possess (3:1) subclustering
(cf.\ Oegerle \& Hill 2001).  
However,
they are more likely to have at least one BCM component with 
a significant peculiar velocity.  We suggest the 
unifying interpretation
of these observations is that our 
dumbbell BCM clusters may be in various states of post-merger
activity.  Those with both subclustering and peculiar velocities
(e.g.\ Abell~3391) are probably in the early stages of mixing:
the bulk of the galaxies are separable in to sub-groups with a
dumbbell component belonging to one (or even each) sub-grouping
(cf.\ Quintana et al.\ 1996).
As the merger progresses, we may 
expect the less massive galaxies to homogenize first 
eventually leaving only a large peculiar velocity 
in one or both dumbbell components as the 
singular signpost to recent cluster activity, before
the individual components of the dumbbell BCM itself relax with 
respect to each other and the rest of the merged cluster (Abell~2911).
Alternatively, we note that intermediate clusters such as 
Abell~3570 may simply be the result
of ellipticals preferentially being on radial orbits 
(e.g.\ Ramirez \& de Souza 1998; see also Hwang \& Lee 2008) 
and hence observationally indistinguishable.  
The dumbbell itself may be a final 
transient phase before ultimately merging into a more 
massive cD-like galaxy (e.g.\ Tremaine 1990; West 1994).  
These results and the emerging picture of
the evolution of dumbbells with respect to cluster substructure are 
also consistent with other observational 
evidence that the timescale for clusters to accrete new galaxies 
is much shorter than the timescale required for the central galaxy 
to merge with the accreted galaxies (e.g. Cooray \& Milosavljevi{\'c} 
2005; Brough et al.\ 2008).

\begin{figure*}[h]
\begin{center}
\vspace*{-1.5in}
\hspace*{-1in}
\includegraphics[scale=1, angle=0, width=8.in]{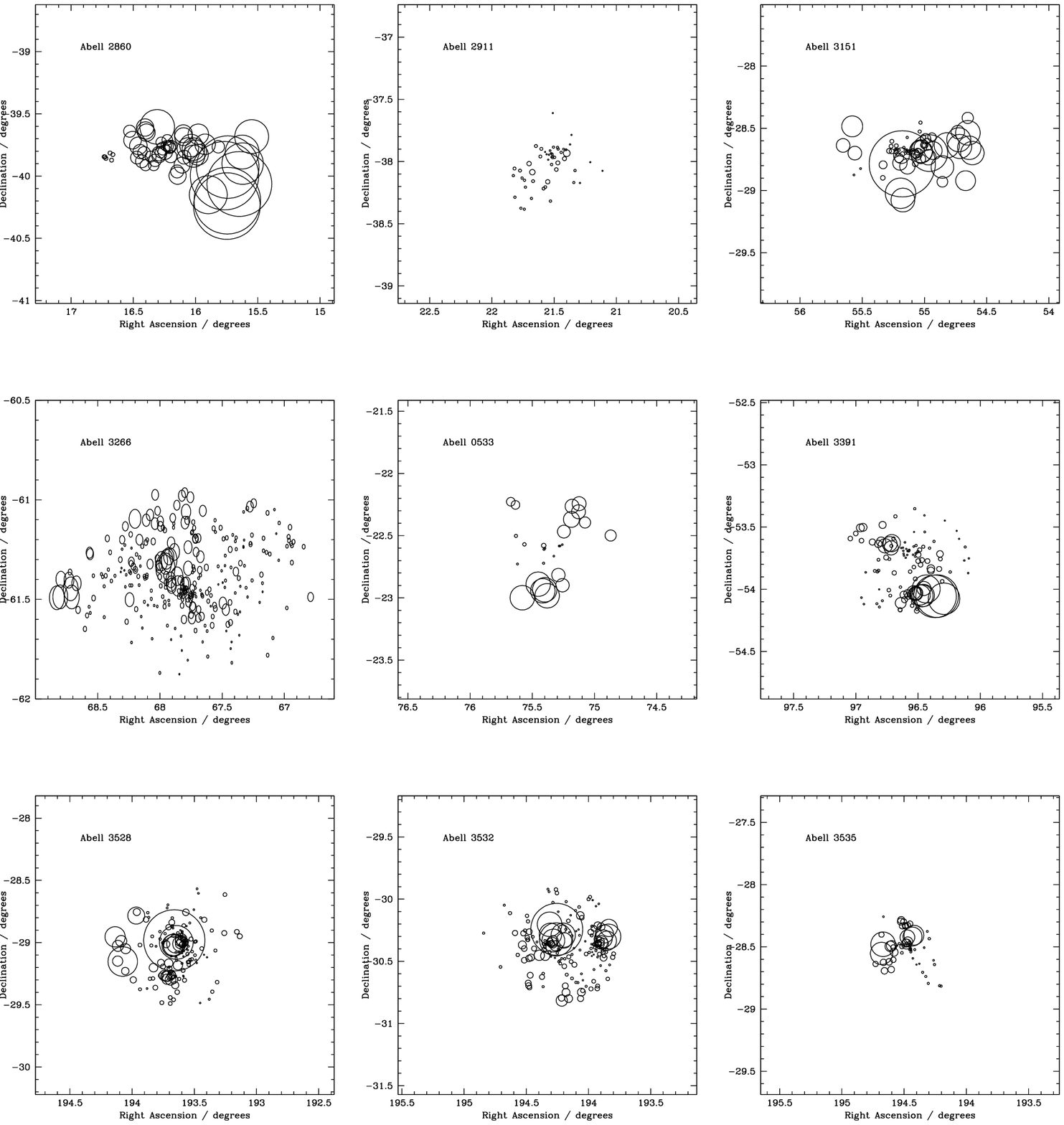}
\vspace*{-1.5in}
\caption{Results of applying the DS test to the dumbbell sample.  
In these bubble plots, each circle's radius is scaled proportional to 
the Dressler \& Shectman (1988) $\delta$ statistic (i.e.\ proportional to
the deviation of a localized group of galaxies 
mean velocity away from the whole cluster's mean velocity).
Substruture is therefore indicated by spatially close and relatively big 
overlapping circles.}\label{fig:multi1}
\end{center}
\end{figure*}

\begin{figure*}[h]
\begin{center}
\vspace*{-1.5in}
\hspace*{-1in}
\includegraphics[scale=1, angle=0, width=8.in]{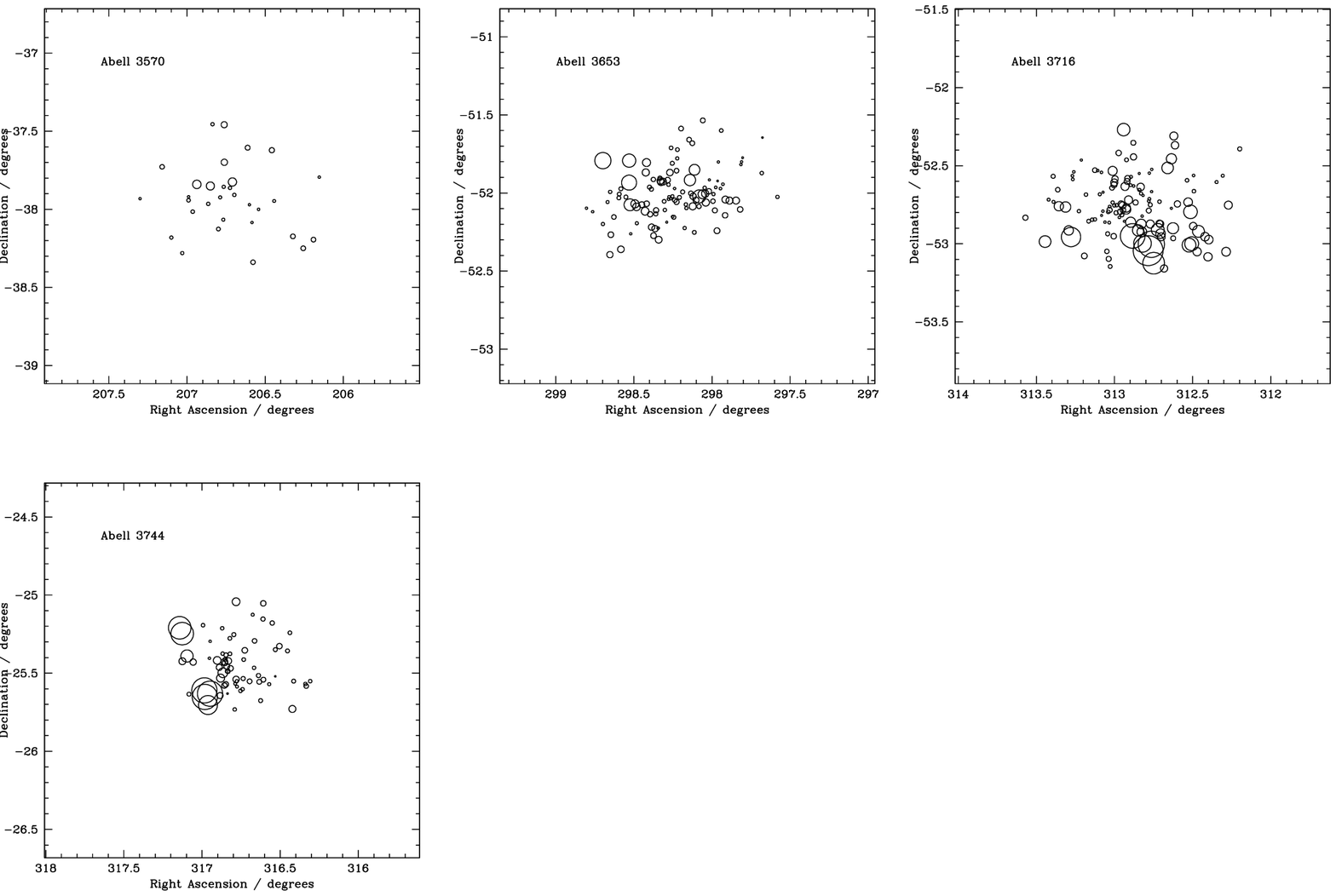}
\vspace*{-4.5in}
\setcounter{figure}{1}
\caption{continued.}\label{fig:multi2}
\end{center}
\end{figure*}

\section*{Acknowledgments} 
We thank the efforts of the 2dFGRS team and Anglo-Australian Observatory
in creating the 2dFGRS survey.
We also thank H.\ Andernach and I.\ G.\ Roseboom for fruitful discussion
during the course of this work and the diligent work of the referee whose
comments have improved this work.
This research has made use of the NASA/IPAC Extragalactic Database 
(NED) which is operated by the Jet Propulsion Laboratory, California 
Institute of Technology, under contract with the National Aeronautics 
and Space Administration.


\end{document}